\documentstyle[12pt,epsfig]{article}                               
\begin{document}

\titlepage                  

\begin{center}
\begin{Large}
\begin{bf}

 Shape-Invariant Single and Double-well Potentials under Spectral Invariance

\vspace{1.0cm}

 Biswanath Rath
\end{bf}
\end{Large}
\end{center}

\vspace{0.1cm}

 Department of Physics,
 Maharaja Sriram Chandra Bhanj Deo University,
 Takatpur, Baripada -757003, Odisha, India.

e.mail:biswanathrath10@gmail.com

\vspace{0.1cm}

\begin{large}

$\bf{Abstract:}$
We formulate  the structure of spectral invariance in shape invariant single and double well potentials using derivative invariance principle.
 
\end{large}

\vspace{1.0cm}

\begin{bf}
\hspace{3.0in} PACS no-03.65.Db
\end{bf}

\vspace{0.1cm}

\begin{bf}
\end{bf}

\vspace{0.1cm}

Supersymmetry [1] has many applications on quantum mechanicalproblems
 involving $V_{\pm}$ reflecting unbroken spectra[2,3,4]. 
 
 Spectral nature becomes interesting, when analyical results are possible.
 
  Mathematically, susy becomes shape-invariance(SI) in nature[5].
   
  Here,we  extend the SI using derivative invariance principle.
\begin{equation}
V_{\pm}=W^{2}-\frac{dW}{dx}
\end{equation}
Here, $W$ is the superpotential. Since, we stress on different models, we change the notion as 
The model susy partner potentials are
\begin{equation}
V^{k}_{\mp}=W_{k}^{2}\mp \frac{dW_{k}}{dx}
\end{equation}
Selection of $k$ is such that 
\begin{equation}
\frac{dW_{1}}{dx}=\frac{dW_{2}}{dx}=....=\frac{dW_{k}}{dx}
\end{equation}
However, wave functions nature are different i.e
\begin{equation} 
\psi_{0}^{(k)}\sim e^{-\int{W_{k} dx}}
\end{equation}
Mathematically,
\begin{equation} 
\psi_{0}^{(1)} \neq \psi_{0}^{(2)}\neq \psi_{0}^{(3)}......
\end{equation}
Here all forms of super potentials $W_{1}$, $W_{2}$ ...$W_{k}$ come  independently  under shapeinvariant in nature.

Let us consder a model example to highlight the above development.

\begin{equation}
W_{1}= A \tanh(x)
\end{equation}

The corresponding SUSY potentials are 
\begin{equation} 
V_{\mp}^{(1)}= A^{2}(\tanh(x))^{2} \mp A (1/\cosh(x))^{2}
\end{equation}

\begin{equation} 
\int{W_{1} dx} = \log_{e} { \cosh(x)}
\end{equation}

\begin{equation}
W_{2}= A \tanh(x + \sinh(\frac{x}{|x|})
\end{equation}

\begin{equation} 
V_{\mp}^{(2)}= A^{2}(\tanh(x + \tanh(\frac{x}{|x|}))^{2} \mp A (1/\cosh(x+\sinh(\frac{x}{|x|}))^{2}
\end{equation}

\begin{equation} 
\int{W_{2} dx} = A \log{(\cosh(x+\sinh(\frac{1}{sgn(x)}))}
\end{equation}

\begin{equation}
W_{3}= A \tanh(x - \tanh(\frac{x}{|x|})
\end{equation}

\begin{equation} 
V_{\mp}^{(3)}= A^{2}(\tanh(x - \tanh(\frac{x}{|x|}))^{2} \mp A(1/\cosh(x-\tanh(\frac{x}{|x|}))^{2}
\end{equation}

\begin{equation} 
\int{W_{3} dx} = A\log{(\cosh(\tanh(\frac{1}{sgn(x)}-x))}
\end{equation}

\begin{equation}
W_{4}= A \tanh(x - \sin(\frac{x}{|x|})
\end{equation}

\begin{equation} 
V_{\mp}^{(4)}= A^{2}(\tanh(x + \sin(\frac{x}{|x|}))^{2} \mp A (1/\cosh(x+\sin(\frac{x}{|x|}))^{2}
\end{equation}

\begin{equation} 
\int{W_{4} dx} =A \log{(\cosh(\sin(\frac{1}{sgn(x)}-x))}
\end{equation}
Now we focus on shape-invariance principle[1-5] i.e
\begin{equation} 
V_{+}=V_{-}+ R(A)
\end{equation}
\begin{equation} 
E_{n}^{-}=\sum R(A)
\end{equation}
More explicitly, we have 
\begin{equation} 
E_{n}^{-}=A^{2}-(A-n)^{2}
\end{equation}

Using computational analysis[6], we find for all the cases, energy levels remain invariant.
in conclusion, suitable choice of superpotential,susy can reflect shapeinvariance principle in single well as well as in double well potentials.
 
\begin{figure}[htbt]
\begin{center}
\includegraphics[height=2.in,width=2.0in]{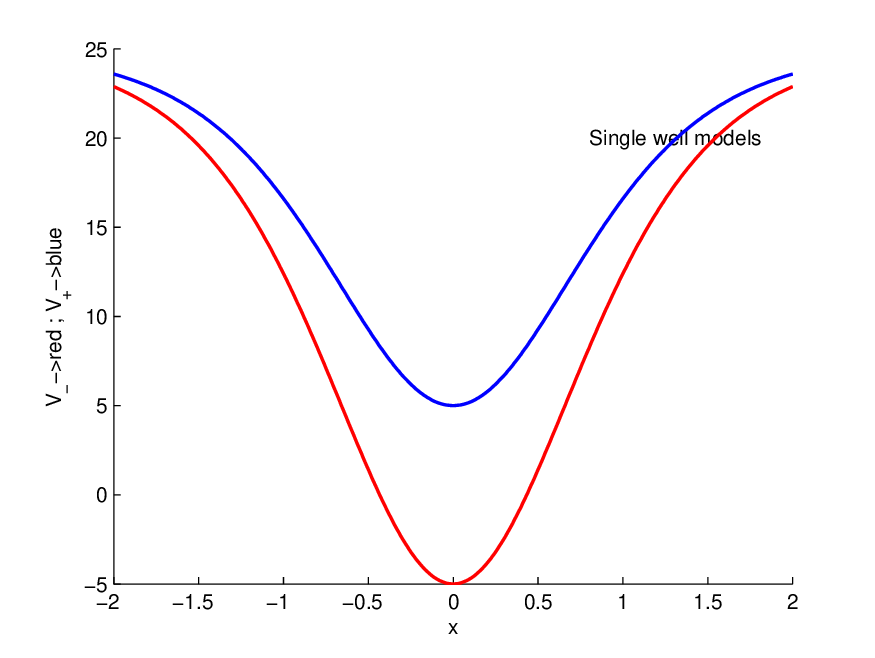}\\
\caption{Single well  }
\end{center}
\end{figure}
\begin{figure}[htbt]
\begin{center}
\includegraphics[height=2.in,width=2.0in]{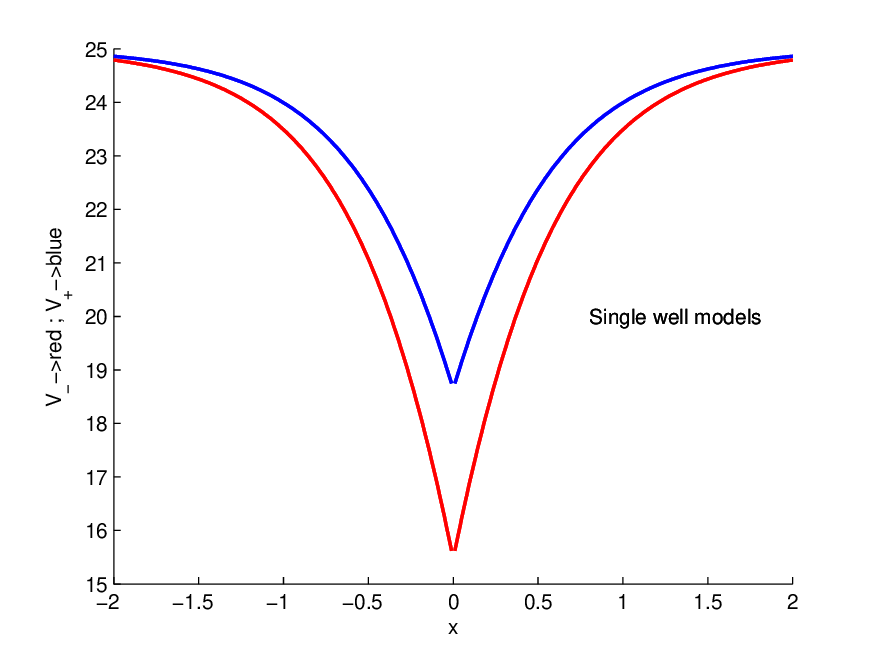}\\
\caption{Single well  }
\end{center}
\end{figure}

\begin{figure}[htbt]
\begin{center}
\includegraphics[height=2.in,width=2.0in]{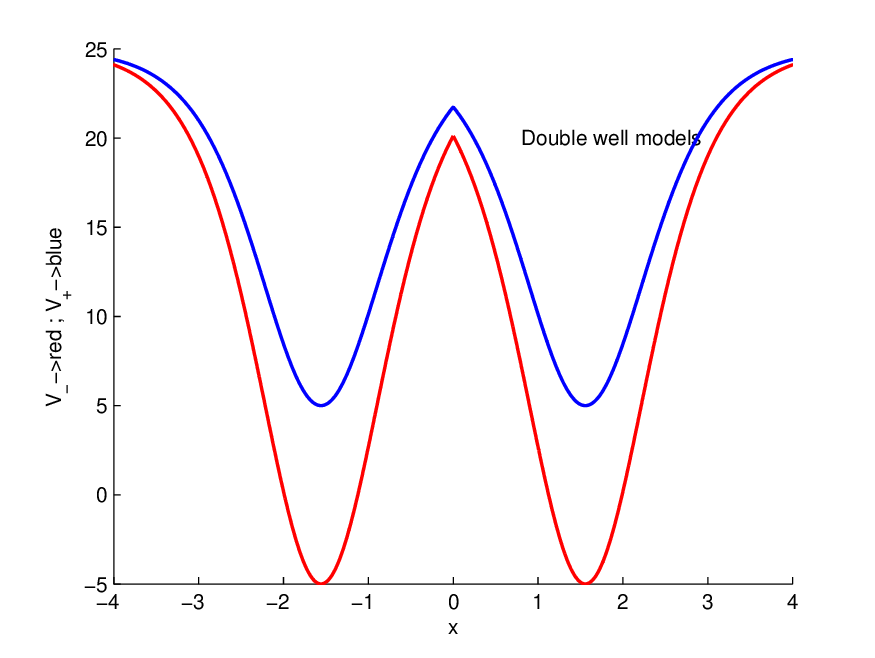}\\
\caption{Double well}
\end{center}
\end{figure}

\begin{figure}[htbt]
\begin{center}
\includegraphics[height=2.in,width=2.0in]{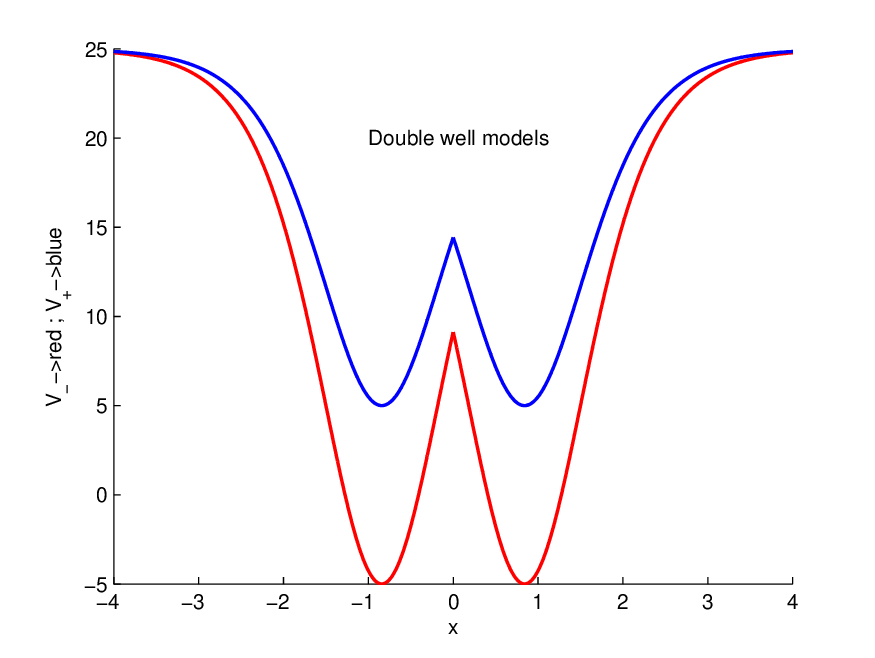}\\
\caption{Double well}
\end{center}
\end{figure}
  
\end{document}